\documentclass[10pt,a4paper,twocolumn]{article}
\usepackage{graphicx}

\begin{document}

\title{The direction of Coulomb's force and the direction of gravitational
force in the 4-dimensional space-time}
\author{Huaiyang Cui \\
Department of physics, Beihang University, Beijing, 100083, China.\\
E-mail: hycui@buaa.edu.cn }
\maketitle

\begin{abstract}
{\small In this paper, we point out that the 4-vector force acting on a
particle is always in the direction orthogonal to the 4-vector velocity of
the particle in the 4-dimensional space-time, rather than along the line
joining the particle and the action source. This inference is obviously
supported from the fact that the magnitude of the 4-vector velocity is a
constant. This orthogonality brings out many new aspects for force concept.
In this paper it is found that the Maxwell's equations can be derived from
classical Coulomb's force and the orthogonality, some gravitational effects
such as the perihelion advance of planet can also be explained in terms of
the orthogonality.}
\end{abstract}
\newline
\newline


\section{ Introduction}

In the world, almost everyone knows that Coulomb's force (or gravitational
force) acts along the line joining a couple of particles, but sometime this
knowledge is incorrect in the theory of relativity. Consider a particle
moving at the 4-vector velocity $u$ in an inertial Cartesian coordinate
system $S:(x_{1},x_{2},x_{3},x_{4}=ict)$, the magnitude of the 4-vector
velocity $u$ is given by\cite{Harris}
\begin{equation}
|u|=\sqrt{u_{\mu }u_{\mu }}=\sqrt{-c^{2}}=ic  \label{1}
\end{equation}
The above equation is valid so that any force can never change $u$ in its
magnitude but can change $u$ in its direction. We therefore conclude that
the Coulomb's force or gravitational force on a particle always acts in the
direction orthogonal to the 4-vector velocity of the particle in the
4-dimensional space-time, rather than along the line joining the particle
and action source. Strictly, any 4-vector force $f$ satisfies the following
orthogonality.
\begin{equation}
u_{\mu }f_{\mu }=u_{\mu }m\frac{du_{\mu }}{d\tau }=\frac{m}{2}\frac{d(u_{\mu
}u_{\mu })}{d\tau }=0  \label{2}
\end{equation}
Where the Cartesian coordinate system $S$ is a frame of reference whose axes
are orthogonal to one another, there is no distinction between covariant and
contravariant components, only subscripts need to be used.

Although the orthogonality has occasionally appeared in some textbooks\cite%
{Rindler}, in the present paper, Eq.(\ref{2}) has been elevated to an
essential requirement for the definition of a force, which brings out many
new aspects for the Coulomb's force and gravitational force.

\section{ The Coulomb's force and Maxwell's equations}

\subsection{ the 4-vector Coulomb's force}

\begin{figure}[tbh]
\includegraphics[bb=175 585 350 730,clip]{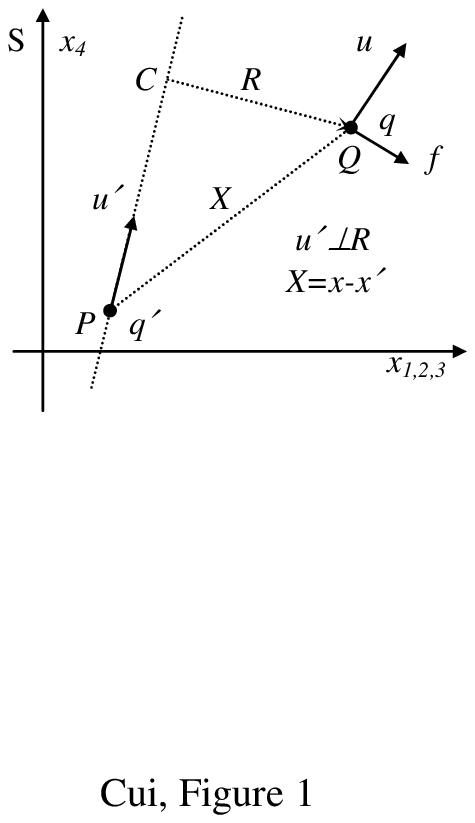}
\caption{The Coulomb's force acting on $q$ is orthogonal to the 4-vector
velocity $u$ of $q$, and lies in the plane of $u^{\prime }$ and $X$ with a
retardation with respect to $q^{\prime }$, here the Euclidian geometry is
used to illustrate the complex space-time.}
\label{fig1}
\end{figure}

Suppose there are two charged particle $q$ and $q^{\prime }$ locating at the
positions $x$ and $x^{\prime }$ respectively in the Cartesian coordinate
system $S$, and moving at the 4-vector velocities $u$ and $u^{\prime }$
respectively, as shown in Fig.\ref{fig1}, where we have used $X$ to denote $%
x-x^{\prime }$. The Coulomb's force $f$ acting on the particle $q$ is
orthogonal to the velocity direction of $q$, as illustrated in Fig.\ref{fig1}
using the Euclidian geometry to represent the complex space-time. Like a
centripetal force, the force $f$ should make an attempt to rotate itself
about the particle path center. We think that the path center, $q$, $%
q^{\prime }$ and the force $f$ all should be in the plane of $u^{\prime }$
and $X$, so that we make an expansion to $f$ as
\begin{equation}
f=Au^{\prime }+BX  \label{3}
\end{equation}
where $A$ and $B$ are unknown coefficients, the possibility of this
expansion will be further clear in the next section in where the expansion
is not an assumption [see Eq.(\ref{25})].

Using the orthogonality $f\perp u$, we get
\begin{equation}
u\cdot f=A(u\cdot u^{\prime })+B(u\cdot X)=0  \label{4}
\end{equation}
By eliminating the coefficient $B$, we rewrite Eq.(\ref{3}) as
\begin{equation}
f=\frac{A}{u\cdot X}[(u\cdot X)u^{\prime }-(u\cdot u^{\prime })X]  \label{5}
\end{equation}
It follows from the direction of Eq.(\ref{5}) that the unit vector $f^{0}$
of the Coulomb's force direction is given by
\begin{equation}
f^{0}=\frac{1}{c^{2}r}[(u\cdot X)u^{\prime }-(u\cdot u^{\prime })X]
\label{6}
\end{equation}
because of $|f^{0}|=1$, where $r=|R|$, $R\perp u^{\prime }$. Suppose that
the magnitude of the force $f$ has the classical form
\begin{equation}
|f|=k\frac{qq^{\prime }}{r^{2}}  \label{9}
\end{equation}
Combination of Eq.(\ref{6}) with Eq.(\ref{9}), we obtain a modified
Coulomb's force
\begin{eqnarray}
f &=&\frac{kqq^{\prime }}{c^{2}r^{3}}[(u\cdot X)u^{\prime }-(u\cdot
u^{\prime })X]  \nonumber \\
&=&\frac{kqq^{\prime }}{c^{2}r^{3}}[(u\cdot R)u^{\prime }-(u\cdot u^{\prime
})R]  \nonumber \\
&=&q[(u\cdot (\frac{kq^{\prime }}{c^{2}r^{3}}R))u^{\prime }-(u\cdot
u^{\prime })(\frac{kq^{\prime }}{c^{2}r^{3}}R)]  \label{10}
\end{eqnarray}
By using the relation
\begin{equation}
\partial _{\mu }(\frac{1}{r})=-\frac{R_{\mu }}{r^{3}}  \label{11}
\end{equation}
we obtain
\begin{eqnarray}
f_{\mu } &=&q[-(u_{\nu }\partial _{\nu }(\frac{kq^{\prime }}{c^{2}r}))u_{\mu
}^{\prime }+(u_{\nu }u_{\nu }^{\prime })\partial _{\mu }(\frac{kq^{\prime }}{%
c^{2}r})]  \nonumber \\
&=&q[-(u_{\nu }\partial _{\nu }(\frac{kq^{\prime }u_{\mu }^{\prime }}{c^{2}r}
))+(u_{\nu })\partial _{\mu }(\frac{kq^{\prime }u_{\nu }^{\prime }}{c^{2}r})]
\label{12}
\end{eqnarray}
The force can be rewritten in terms of 4-vector components as
\begin{eqnarray}
f_{\mu } &=&qF_{\mu \nu }u_{\nu }  \label{13} \\
F_{\mu \nu } &=&\partial _{\mu }A_{\nu }-\partial _{\nu }A_{\mu }  \label{14}
\\
A_{\mu } &=&\frac{kq^{\prime }u_{\mu }^{\prime }}{c^{2}r}  \label{15}
\end{eqnarray}
Thus $A_{\mu }$ expresses the 4-vector potential of the particle $q^{\prime
} $. It is easy to find that Eq.(\ref{13}) contains the Lorentz force.

\subsection{ The Lorentz gauge condition and the Maxwell's equations}

From Eq.(\ref{15}), because of $u^{\prime }\perp R$ , i.e. $u_{\mu }^{\prime
}R_{\mu }=0$, we have
\begin{equation}
\partial _{\mu }A_{\mu }=\frac{kq^{\prime }u_{\mu }^{\prime }}{c^{2}}
\partial _{\mu }(\frac{1}{r})=-\frac{kq^{\prime }u_{\mu }^{\prime }}{c^{2}}(%
\frac{R_{\mu }}{r^{3}})=0  \label{16}
\end{equation}
It is known as the Lorentz gauge condition.

To note that $R$ has three degrees of freedom under the condition of $R\perp
u^{\prime }$, so we have
\begin{eqnarray}
\partial _{\mu }R_{\mu } &=&3  \label{17} \\
\partial _{\mu }\partial _{\mu }(\frac{1}{r}) &=&-4\pi \delta (R)  \label{18}
\end{eqnarray}
From Eq.(\ref{14}), we have
\begin{eqnarray}
\partial _{\nu }F_{\mu \nu } &=&\partial _{\nu }\partial _{\mu }A_{\nu
}-\partial _{\nu }\partial _{\nu }A_{\mu }=-\partial _{\nu }\partial _{\nu
}A_{\mu }  \nonumber \\
&=&-\frac{kq^{\prime }u_{\mu }^{\prime }}{c^{2}}\partial _{\nu }\partial
_{\nu }(\frac{1}{r})  \nonumber \\
&=&\frac{kq^{\prime }u_{\mu }^{\prime }}{c^{2}}4\pi \delta (R)=\mu
_{0}J_{\nu }^{\prime }  \label{19}
\end{eqnarray}
where we define $J_{\nu }^{\prime }=q^{\prime }u_{\nu }^{\prime }\delta (R)$
as the current density of the source. From Eq.(\ref{14}), by exchanging the
indices and taking the summation of them, we have
\begin{equation}
\partial _{\lambda }F_{\mu \nu }+\partial _{\mu }F_{\nu \lambda }+\partial
_{\nu }F_{\lambda \mu }=0  \label{20}
\end{equation}
The Eq.(\ref{19}) and (\ref{20}) are known as the Maxwell's equations. For
continuous media, they are valid as well as.

\subsection{ The Lienard-Wiechert potential}

From the Maxwell's equations, we know there is a retardation time for the
electromagnetic action to propagate between the two particles, as
illustrated in Fig.\ref{fig1}, the retardation effect is measured by the
distance from point P to the point C.
\begin{equation}
r=c\Delta t=c\frac{|PC|}{ic}=c\frac{u^{\prime }{}^{0}\cdot X}{ic}=\frac{%
u_{\nu }^{\prime }(x_{\nu }^{\prime }-x_{\nu })}{c}  \label{21}
\end{equation}
Then
\begin{equation}
A_{\mu }=\frac{kq^{\prime }}{c^{2}}\frac{u_{\mu }^{\prime }}{r}=\frac{%
kq^{\prime }}{c}\frac{u_{\mu }^{\prime }}{u_{\nu }^{\prime }(x_{\nu
}^{\prime }-x_{\nu })}  \label{22}
\end{equation}
Obviously, Eq.(\ref{22}) is known as the Lienard-Wiechert potential for a
moving particle.

The above formalism clearly shows that the Maxwell's equations can be
derived from the classical Coulomb's force and the orthogonality of 4-vector
force and 4-vector velocity. In other words, the orthogonality is hidden in
the Maxwell's equation. Specially, Eq.(\ref{5}) directly accounts for the
geometrical meanings of the curl of vector potential, the curl contains the
orthogonality. The orthogonality of force and velocity is one of
consequences from the relativistic mechanics.

\section{ Gravitational force}

\subsection{ the 4-vector gravitational force}

The above formalism has a significance on guiding how to develop the theory
of gravity. In analogy with the modified Coulomb's force of Eq.(\ref{10}),
we directly suggest a modified universal gravitational force as
\begin{eqnarray}
f &=&-\frac{Gmm^{\prime }}{c^{2}r^{3}}[(u\cdot X)u^{\prime }-(u\cdot
u^{\prime })X]  \nonumber \\
&=&-\frac{Gmm^{\prime }}{c^{2}r^{3}}[(u\cdot R)u^{\prime }-(u\cdot u^{\prime
})R]  \label{23}
\end{eqnarray}
for a couple of particles with masses $m$ and $m^{\prime }$, respectively,
the gravitational force must satisfy the orthogonality of 4-vector force and
4-vector velocity.

Even though the gravitation is treated as a field in the theory of general
relativity, here we emphasize that gravitational force must hold the
orthogonality if it can be defined as a force. It follows from Eq.(\ref{23})
that we can predict that there exist the gravitational radiation and
magnetism-like components for a gravitational force. Particularly, the
magnetism-like components would play an important role in geophysics and
atmosphere physics.

If we have not any knowledge but know there exists the classical universal
gravitation $\mathbf{f}$ between two particles $m$ and $m^{\prime }$, what
form will take the 4-vector gravitational force $f$ ? Suppose that $%
m^{\prime }$ is at rest at the origin, using $u=(\mathbf{u},u_{4})$, $%
u^{\prime }=(0,0,0,ic)$ and $u\cdot f=0$, we have
\begin{eqnarray}
f_{4} &=&\frac{u_{4}f_{4}}{u_{4}}=\frac{u\cdot f-\mathbf{u}\cdot \mathbf{f}}{%
u_{4}}=-\frac{\mathbf{u}\cdot \mathbf{f}}{u_{4}}  \label{24} \\
f &=&(\mathbf{f},f_{4})=(\mathbf{f},0)+(0,0,0,f_{4})  \nonumber \\
&=&(\mathbf{f},0)+f_{4}\frac{(0,0,0,ic)}{ic}  \nonumber \\
&=&\frac{|\mathbf{f}|}{|\mathbf{R}|}(\mathbf{R},0)+f_{4}\frac{u^{\prime }}{ic%
}=\frac{|\mathbf{f}|}{|\mathbf{R}|}(\mathbf{R},0)-(\frac{\mathbf{u}\cdot
\mathbf{f}}{u_{4}})\frac{u^{\prime }}{ic}  \nonumber \\
&=&\frac{|\mathbf{f}|}{|\mathbf{R}|}\frac{1}{icu_{4}}[icu_{4}(\mathbf{R},0)-(%
\mathbf{u}\cdot \mathbf{R})u^{\prime }]  \nonumber \\
&=&\frac{|\mathbf{f}|}{|\mathbf{R}|}\frac{1}{(u^{\prime }\cdot u)}%
[(u^{\prime }\cdot u)(\mathbf{R},0)-(\mathbf{u}\cdot \mathbf{R})u^{\prime }]
\nonumber \\
&=&\frac{|\mathbf{f}|}{|\mathbf{R}|(u^{\prime }\cdot u)}[(u^{\prime }\cdot
u)R-(u\cdot R)u^{\prime }]  \label{25}
\end{eqnarray}%
where $R\perp u^{\prime }$, $R=(\mathbf{R},0),r=|\mathbf{R}|$. If we
\textquotedblright rotate\textquotedblright\ (the Lorentz transformations)
our frame of reference in the 4-dimensional space-time to make $m^{\prime }$
not to be at rest, Eq.(\ref{25}) will still valid because of its covariance.
Then we find the 4-vector gravitational force goes back to the form of Eq.(%
\ref{23}), like Lorentz force, having magnetism-like components.

Consider a planet $m$ moving about the sun $M$ with the 4-vector velocity $%
u=(\mathbf{u},u_{4})$, $\mathbf{u}=\mathbf{u}_{r}+\mathbf{u}_{\varphi }$ in
the polar coordinate system $S:(r,\varphi ,ict)$. We assume the sun is at
rest at the origin, $u^{\prime }=(0,0,0,ic)$, then from Eq.(\ref{23}), the
motion of the planet is governed by
\begin{eqnarray}
\frac{du_{4}}{d\tau } &=&-\frac{GM}{c^{2}r^{3}}(ru_{r})ic  \label{29} \\
\frac{d\mathbf{u}}{d\tau } &=&\frac{GM}{c^{2}r^{3}}(icu_{4})\mathbf{r}
\label{30}
\end{eqnarray}%
From the above equations we obtain their solutions
\begin{eqnarray}
u_{4} &=&ic(\varepsilon +\frac{s}{2r})  \label{31} \\
u_{r}^{2}+u_{\varphi }^{2}+u_{4}^{2} &=&-c^{2}  \label{32} \\
ru_{\varphi } &=&h  \label{33}
\end{eqnarray}%
where $\varepsilon $ and $h$ are two integral constants, $s=2GM/c^{2}$ is
called the Schwarzschild's radius, $s/r$ is a small term. Eq.(\ref{32})
agrees with the relation $u^{2}=-c^{2}$. Consider the turning point at where
the planet is at rest, $u_{4}=ic$, the radius of the turning point is $%
r_{tp} $, then

\begin{equation}
\varepsilon =1-\frac{s}{2r_{tp}}  \label{331}
\end{equation}

Eq.(\ref{31}) tells us the relation between the time interval $dt$ and the
proper time interval $d\tau $, i.e.
\begin{equation}
u_{4}=\frac{icdt}{d\tau }=ic(\varepsilon +\frac{s}{2r})  \label{34}
\end{equation}%
Hence, we know that the proper time interval $d\tau $ at position $r$ and
the proper time interval $d\tau _{\infty }$ at $r=\infty $ have the relation%
\begin{equation}
\frac{d\tau }{d\tau _{\infty }}=\frac{ic\varepsilon }{ic(\varepsilon +s/(2r))%
}\approx 1-\frac{s}{2r}  \label{35}
\end{equation}%
Since $s/r$ (or $s/(2h^{2}/c^{2})$) is a very small quantity, we neglect all
but the first-order terms.

In the following subsections, we show that Eq.(\ref{31}), (\ref{32}) and (%
\ref{33}) can give out the same results as the theory of general relativity
for gravitational problems\cite{Cui}.

\subsection{ The gravitational red shift}

According to Eq.(\ref{35}), the period of oscillation of an atom at $r$ is
related to the period of oscillation of an atom at $r=\infty $ by
\begin{equation}
\frac{T_{r}}{T_{\infty }}\approx 1-\frac{s}{2r}  \label{36}
\end{equation}%
this effect is called the gravitational red shift\cite{Foster}.

\subsection{ The perihelion advance of a planet}

Consider the planet again, Substituting Eq.(\ref{35}) into Eq.(\ref{32}) and
Eq.(\ref{33}), we obtain
\begin{eqnarray}
\lbrack (1+\frac{s}{2r})\frac{dr}{d\tau _{\infty }}]^{2}+[(1+\frac{s}{2r})%
\frac{dx_{\varphi }}{d\tau _{\infty }}]^{2}+[ic(\varepsilon +\frac{s}{2r}%
)]^{2} \nonumber \\
=-c^{2} \label{37}\\
r(1+\frac{s}{2r})\frac{dx_{\varphi }}{d\tau _{\infty }} =h
\label{38}
\end{eqnarray}%
They also can be rewritten as
\begin{eqnarray}
(\frac{dr}{d\tau _{\infty }})^{2}+\frac{h^{2}}{r^{2}}(1-\frac{s}{r}%
)-c^{2}[(\varepsilon ^{2}-1)+\varepsilon
\frac{s}{r})](1-\frac{s}{r}) \nonumber \\
=0 \label{39} \\
r(1+\frac{s}{2r})\frac{dx_{\varphi }}{d\tau _{\infty }} =h
\label{40}
\end{eqnarray}%
In order to pursue the covariance of Eq.(\ref{40}), we define the angular
displacement $d\varphi _{\infty }=[1+s/(2r)]d\varphi $, then we have
\begin{eqnarray}
(\frac{dr}{d\tau _{\infty }})^{2}+\frac{h^{2}}{r^{2}}(1-\frac{s}{r}%
)-(\varepsilon -\varepsilon ^{2}+1)\frac{sc^{2}}{r} \nonumber \\
=(\varepsilon ^{2}-1)c^{2} \label{41} \\
r^{2}\frac{d\varphi _{\infty }}{d\tau _{\infty }} =h  \label{42}
\end{eqnarray}%
Eliminating $d\tau _{\infty }$ by dividing Eq.(\ref{41}) by Eq.(\ref{42}),
we obtain
\begin{eqnarray}
(\frac{dr}{d\varphi _{\infty
}})^{2}+r^{2}(1-\frac{s}{r})-(\varepsilon -\varepsilon
^{2}+1)\frac{sc^{2}}{h^{2}}r^{3} \nonumber \\
=(\varepsilon^{2}-1)\frac{ c^{2}}{h^{2}}r^{4} &&  \label{43}
\end{eqnarray}%
Now making the change of variable $U=1/r$ gives
\begin{eqnarray}
(\frac{dU}{d\varphi _{\infty }})^{2}+U^{2}(1-sU)-(\varepsilon
-\varepsilon ^{2}+1)\frac{sc^{2}}{h^{2}}U \nonumber\\
=(\varepsilon ^{2}-1)\frac{c^{2}}{h^{2}} \label{44}
\end{eqnarray}%
Differentiating with respect to $\varphi _{\infty }$ gives
\begin{equation}
\frac{d^{2}U}{d\varphi _{\infty }^{2}}+U-\frac{3s}{2}U^{2}=(\varepsilon
-\varepsilon ^{2}+1)\frac{sc^{2}}{2h^{2}}  \label{45}
\end{equation}%
The last term on the left side of the equation is relativistic correction,
in the absence of this term, the solution is
\begin{eqnarray}
U &=&(\varepsilon -\varepsilon ^{2}+1)\frac{c^{2}s}{2h^{2}}[1+e\cos (\varphi
_{\infty }-\varphi _{0})]  \nonumber \\
&=&U_{0}[1+e\cos (\varphi _{\infty }-\varphi _{0})]  \label{47}
\end{eqnarray}%
where $e$ and $\varphi _{0}$ are the constants of integration. Writing $%
U=U_{0}+U_{1}$, we find that Eq.(\ref{45}) becomes
\begin{equation}
\frac{d^{2}U_{1}}{d\varphi _{\infty }^{2}}+(1-3sU_{0})U_{1}=\frac{3s}{2}%
(U_{0}^{2}+U_{1}^{2})  \label{48}
\end{equation}%
By inspection we see that $U_{1}$ is on oscillatory function of $\varphi
_{\infty }$ with the frequency
\begin{equation}
\omega =\sqrt{1-3sU_{0}}  \label{49}
\end{equation}%
From $\omega \phi =2\pi $ for a cycle, we know that
\begin{equation}
\phi \approx 2\pi (1+\frac{3}{2}sU_{0})  \label{50}
\end{equation}%
Thus the perihelion advance of the planet is given by
\begin{equation}
\Delta \phi =\phi -2\pi =3\pi sU_{0}  \label{51}
\end{equation}%
This result is the same as that in the theory of general relativity.

\subsection{ The bending of light rays}

Since the above solutions are independent from the planet mass $m$, if the
planet is a photon without mass, it also satisfies the above motion like a
planet. For a photon, $d\tau =0$, Eq.(\ref{33}) becomes
\begin{equation}
h=r^{2}\frac{d\varphi }{d\tau }=\infty  \label{52}
\end{equation}%
Then Eq.(\ref{45}) becomes
\begin{equation}
\frac{d^{2}U}{d\varphi _{\infty }^{2}}+U-\frac{3}{2}sU^{2}=0  \label{53}
\end{equation}%
The solution of this equation gives out the path of the photon, it is the
same as that of the theory of general relativity. So in its flight past the
massive object $M$ of radius $R$ the photon is deflected through an angle

\begin{equation}
\delta =\frac{2s}{R}  \label{54}
\end{equation}%
in agreement with experimental observations.

\subsection{ The bending of space}

If there is not the nonlinear effect of photon flying, any straight line in
the space-time is of the path of photon. Since the light is bent in the
4-dimensional space-time for an infinite distance observer, we consequently
conclude that the space will bend due to the gravitational effect mentioned
above near a massive object.

\subsection{ The Lorentz transformation}

In classical mechanics, we have the impression that force yields
acceleration, here we make an attempt to express that force yields the
rotation of 4-vector velocity in the 4-dimensional space time.

When a particle is accelerated from a rest state at the $r$ position to a
speed $v$ state at the $r^{\prime }$ position by a gravitational field $\Phi
$, the 4-th axis of the frame $S$ fixed at the particle will rotates an
angle $\theta $ (towards the direction of $v$), the frame $S$ will become a
new frame $S^{\prime }$ in the viewpoint on the ground, the angle is given
by $u_{4}=ic\cosh \theta $. From Eq.(\ref{31}), we get%
\begin{eqnarray}
u_{4}|_{r} &=&ic=ic(\varepsilon +\frac{s}{2r})  \label{55} \\
u_{4}|_{r^{\prime }} &=&ic\cosh \theta =ic(\varepsilon +\frac{s}{2r^{\prime }%
})  \label{56} \\
\cosh \theta &=&\frac{u_{4}|_{r^{\prime }}}{u_{4}|_{r}}=1+\frac{s}{%
2\varepsilon r^{\prime }}-\frac{s}{2\varepsilon r}\approx 1+\frac{\Phi }{%
c^{2}}  \label{57}
\end{eqnarray}%
where $\Phi =c^{2}s/(2r^{\prime })-c^{2}s/(2r)$ is the potential difference
between $S$ and $S^{\prime }$. Since the particle is accelerated by $\Phi $,
then the rotation is given by
\begin{eqnarray}
\frac{1}{2}v^{2} &=&\Phi \qquad  \label{58} \\
\cosh \theta &\approx &1+\frac{\Phi }{c^{2}}\approx \frac{1}{\sqrt{%
1-v^{2}/c^{2}}}  \label{59} \\
\sinh \theta &=&\sqrt{1-\cosh ^{2}\theta }=\frac{iv/c}{\sqrt{1-v^{2}/c^{2}}}
\label{60}
\end{eqnarray}%
Because the origins of $S$ and $S^{\prime }$ coincide at the time $t=0$, the
coordinate transformations between them are
\begin{eqnarray}
dr &=&dr^{\prime }\cosh \theta +dx_{4}^{\prime }\sinh \theta  \nonumber \\
&=&\frac{1}{\sqrt{1-v^{2}/c^{2}}}dr^{\prime }+\frac{iv/c}{\sqrt{1-v^{2}/c^{2}%
}}dx_{4}^{\prime }  \label{61} \\
dx_{4} &=&-dr^{\prime }\sinh \theta +dx_{4}^{\prime }\cosh \theta  \nonumber
\\
&=&\frac{-iv/c}{\sqrt{1-v^{2}/c^{2}}}dr^{\prime }+\frac{1}{\sqrt{%
1-v^{2}/c^{2}}}dx_{4}^{\prime }  \label{62}
\end{eqnarray}%
It is known as the Lorentz transformations.

\section{ Discussion}

To note that the orthogonality of 4-vector force and 4-vector velocity is
valid for any force: strong, electromagnetic, weak and gravitational
interactions, therefore many new aspects of force concept in mechanics
remain for physics to explore.

\section{ Conclusions}

This paper points out that the 4-vector force acting on a particle is always
in the direction orthogonal to the 4-vector velocity of the particle in the
4-dimensional space-time, rather than along the line joining the particle
and the action source. This inference is obviously supported from the fact
that the magnitude of the 4-vector velocity is a constant. This
orthogonality brings out many new aspects for force concept. In this paper
it is found that the Maxwell's equations can be derived from classical
Coulomb's force and the orthogonality, some gravitational effects such as
the perihelion advance of planet can also be explained in terms of the
orthogonality.

\end{document}